\documentclass[%
fleqn,
aps,
twocolumn,
groupedaddress,
prb,
amsmath,amssymb,
floatfix,
10pt
]{revtex4-1}

\usepackage{graphicx}
\usepackage{dcolumn}
\usepackage{bm}

\usepackage{float}
\usepackage[font=footnotesize, labelfont={sf,bf}, skip=0pt
]{caption}
\usepackage[skip=0pt]{subcaption}

\usepackage{ mathrsfs } 				
\usepackage{color}
\usepackage[colorlinks=true, allcolors=blue]{hyperref}
\usepackage{xifthen}					
\usepackage{xparse}					
\usepackage{cancel}
\usepackage{bbold}					
\usepackage{url}

\newcommand{\bra}[2][]{
\ifthenelse{\isempty{#1}}								
{\Big\langle #2 \Big|}									
{#1\langle #2 #1|}
}

\newcommand{\ket}[2][]{
\ifthenelse{\isempty{#1}}								
{\Big| #2 \Big\rangle}									
{#1| #2 #1\rangle}
}

\newcommand{\projection}[3][]{
\ifthenelse{\isempty{#1}}								
{\Big\langle #2\Big|#3 \Big\rangle}						
{#1\langle #2 #1| #3 #1\rangle}
}

\newcommand{\projectO}[2][]{
\ifthenelse{\isempty{#1}}								
{\Big| #2\Big\rangle\Big\langle#2\Big|}					
{#1| #2 #1\rangle #1\langle #2 #1|}
}

\newcommand{\transition}[4][]{
\ifthenelse{\isempty{#1}}								
{\Big\langle #2\Big|#3\Big|#4\Big\rangle}					
{#1\langle #2 #1|#3 #1|#4 #1\rangle}
}

\newcommand{\expect}[2][]{
\ifthenelse{\isempty{#1}}								
{\Big \langle #2\Big\rangle}							
{#1 \langle #2 #1\rangle}								
}

\DeclareDocumentCommand{\trace}{O{} O{} m}{
\ifthenelse{\isempty{#1}}								
{													
\ifthenelse{\isempty{#2}}								
{\mathrm{tr}\Big( #3\Big)}								
{\mathrm{tr}_{#2}\Big( #3\Big)}							
}
{													
\ifthenelse{\isempty{#2}}								
{\mathrm{tr} #1( #3 #1)}								
{\mathrm{tr}_{#2} #1( #3 #1)}							
}
}

\newcommand{\pd}[3][]{%
\ifthenelse{\isempty{#1}}{{\partial #2\over \partial #3}}{{\partial^{#1} #2\over \partial #3^{#1}}}%
}
\newcommand{\pdp}[4][]{
\ifthenelse{\isempty{#1}}{{\left.\pd{#2}{#3}\right|_{#3=#4}}}{{\left.\pd[#1]{#2}{#3}\right|_{#3=#4}}}	
}
\newcommand{\D}[3][]{%
\ifthenelse{\isempty{#1}}{{d #2\over d #3}}{{d^{#1} #2\over d #3^{#1}}}%
}
\newcommand{\Dp}[4][]{
\ifthenelse{\isempty{#1}}{{\left.\D{#2}{#3}\right|_{#3=#4}}}{{\left.\D[#1]{#2}{#3}\right|_{#3=#4}}}	
}

\newcommand{\beq}{\begin{equation}}
\newcommand{\eeq}{\end{equation}}

\newcommand{\dinger}{Schr\"{o}dinger }

\newcommand{\Ham}{\mathcal{H}}

\newcommand{\e}{\varepsilon}

\newcommand{\om}{\omega}
\newcommand{\si}{\sigma}

\newcommand{\half}{\frac{1}{2}}
\newcommand{\up}{\uparrow}
\newcommand{\down}{\downarrow}

\begin{document}

\preprint{APS/123-QED}

\title{Non-Monotonic Thermoelectric Currents and Energy Harvesting in 
Interacting Double Quantum-Dots}

\author{Yair Mazal}
\affiliation{Department of Physics and the Ilse Katz Center for nanoscale 
Science and Technology, Ben-Gurion University of the Negev, Beer Sheva, Israel}
\author{Yigal Meir}
\affiliation{Department of Physics and the Ilse Katz Center for nanoscale 
Science and Technology, Ben-Gurion University of the Negev, Beer Sheva, Israel}
\author{Yonatan Dubi}
\affiliation{Department of Chemistry and the Ilse Katz Center for nanoscale 
Science and Technology, Ben-Gurion University of the Negev, Beer Sheva, Israel}

\date{\today}

\begin{abstract}
A theoretical study of the thermoelectric current and energy harvesting in an 
interacting double quantum dot system, connected to reservoirs held at different 
chemical potentials and temperatures is presented. Using a rate-equation 
approach, the current is evaluated for different energetic configurations of 
the double quantum dot. We discuss in detail the current-temperature gradient 
relations (the thermoelectric analog to current-voltage relations), and 
demonstrate that, due to interactions, the current is non-monotonically 
dependent on thermal bias. This interaction-induced non-monotonicity influences 
the possibility of harvesting thermal energy from the double-quantum dot, and 
it is shown that in some configurations, energy cannot be harvested at all.  
We analyze  the conditions under which energy can indeed be harvested and 
converted to useful electrical power, and the optimal conditions for 
thermoelectric energy conversion.

\end{abstract}

\pacs{Valid PACS appear here}
\maketitle

\section{\label{sec:intro}Introduction}
In bulk systems, the efficiency of thermoelectric harvesting - generation of an 
electrical current and voltage from a temperature gradient - is limited by the 
material properties. The prospect of using nano-scale systems to enhance 
thermoelectric performance \cite{Mahan1996,Dresselhaus1999,Dresselhaus2007}, 
along with the huge advances in constructing and manipulating nanoscale 
structures and devices, now make "nanoscale thermoelectrics" a large and 
interdisciplinary field of research. Indeed, nanoscale thermoelectric energy 
harvesting has now been demonstrated in systems such as nano-composite 
materials, carbon and silicon nanowires, Graphene nanostructures, 
molecular junctions, naoparticles and quantum dots 
(see, e.g., Refs. \onlinecite{wang2014,cui2017,benenti2017fundamental} and references therein). 

Besides the obvious applicative interest, in recent years it was realized that 
studying thermoelectric effects in nanoscale systems can shed light on the 
transport mechanisms dominating nanoscale structures. For instance, in 
molecular junctions, the thermoelectric voltage and thermopower can distinguish 
between different transport mechanisms \cite{Segal2005,Li2016,Korol2016}. 
Thermopower in quantum dots was studied over two decades ago, both 
theoretically and experimentally \cite{Beenakker1992,Staring1993,thierschmann2016thermoelectrics}, with recent 
resurging interest focusing on role of interactions and non-linearity on 
thermo-electric efficiency  
\cite{Erdman2017,Whitney2014,Talbo2017,Josefsson2018,Kennes2013,Sanchez2016,Thierschmann2015,Svilans2016,Svensson2013,Vicioso2018} 
(see review in, e.g., Ref.~\onlinecite{Sothmann2015}).

Double quantum dots (DQDs) are an excellent platform for studying the interplay 
between interactions, quantum effects (i.e. interference), charge and heat 
transport and thermoelectricity \cite{Sierra2016,Gong2012,Chi2011,Rajput2011,Costi2010,Tsaousidou2010,Tagani2012,Wojcik2016,Donsa2014,Zhang2007,Wierzbicki2011,Chen2000,Trocha2012,VanderWiel2002}. 
The possibility of arranging them in parallel or series, to couple them either 
by tunneling or capacitively, and to tune each quantum dot energy separately, 
gives rise to a large spectrum of parameters which can be tuned, leading to a 
broad spectrum of thermoelectric phenomena. Here, we study the effect of DQD 
parameters (level spacing, interaction strength etc.) on thermoelectric energy 
harvesting. We focus on the non-linear response regime, with finite voltages and finite 
temperature difference. We show that at given temperature difference and finite 
bias, one cannot always harvest the thermal bias into electric power, 
and find the conditions for energy harvesting under different DQD parameters. 
We show that, surprisingly, in the presence of  strong interactions such that transport is dominated by a single channel, there is a 
minimal temperature difference under which no harvesting is possible, and 
provide the mechanism for this effect. 

The paper is organized as follows. In Sec.~\ref{formalism} the method and model 
are detailed. In Sec.~\ref{current} we discuss the dependence of current on 
temperature difference in the non-linear regime. In Sec.~\ref{harvesting} we 
discuss the conditions for energy harvesting, and Sec.~\ref{summary} is devoted 
to summary and conclusions.

\section{Formalism} \label{formalism}

\subsection{Model}
\begin{figure}
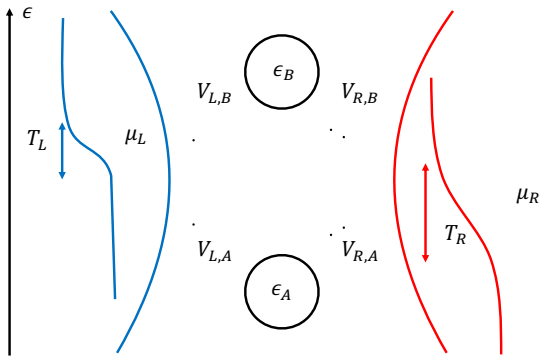

	\centering
			\includegraphics[width=0.5\textwidth]{{{fig_1_sys_illus}}}
	\caption[]{
		\label{fig:sys_illus}%
		An illustration of the system. We consider two levels with level spacing 
		$\Delta\epsilon=\epsilon_B-\epsilon_A$. Each is connected to two leads 
		($V_{x,i}$), with possibly different temperatures $T_{L(R)}$ and 
		chemical potentials $\mu_{L(R)}$.
	}
\end{figure}

The system under consideration is a DQD,  illustrated in Fig. 
\ref{fig:sys_illus}. Within the DQD we take into account intra-dot as well as 
inter-dot Coulomb interaction. The DQD is coupled to two leads characterized 
by different temperatures and chemical potentials.
The corresponding Hamiltonian is:
\begin{align}
\label{eq:Main_Hamiltonian}
\Ham &=\sum_{\substack{k,\si\\ x=L,R}}\e_{kx}c_{kx\si}^\dagger c_{kx\si}
+\Ham_{DQD}+\\
&+\sum_{\substack{k\si\\ i=A,B\\ x=L,R}} V_{kx i\si}c_{kx\si}^\dagger 
d_{i\si} +h.c. ,\nonumber
\end{align}
where $\e_{kx}$ is the energy of an electron with momentum $k$ in the $x$-th 
lead ($x=L,R$), $c_{kx\si}^\dagger$ creates an electron with spin $\si$ and 
momentum $k$ in the lead, $d_{i\si}$ annihilates an electron with spin $\si$ 
in the $i$-th dot, and $V_{kx i\si}$ is the coupling between the lead and the 
dot. The first term is the Hamiltonian of the leads,  the second describes the 
DQD,  and the third represents the interface. 
The Hamiltonian of the DQD can be written as:
\begin{equation}
\label{eq:DQD_Ham}
\Ham_{DQD}=\sum_{\substack{\si,\\ i=A,B}}\e_i d_{i\si}^\dagger d_{i\si}
+U_0 \sum_{i=A,B}\hat{n}_{i\down}\hat{n}_{i\up}+U_1 \hat{n}_A\hat{n}_B  ~,
\end{equation}
where $\e_i$ is the level of the $i$-th dot, $U_0 (U_1)$ is the inter-dot 
(intra-dot) Coulomb interaction strength, 
$\hat{n}_{i\sigma}\equiv d_{i\si}^\dagger d_{i\si}$, and $\hat{n}_i=
\hat{n}_{i\down}+\hat{n}_{i\up}$. 
All energies are measured from the zero-energy, defined as the energy of the 
empty dot. For simplicity, and to reduce the number of numerical parameters, we 
assume that $U_0=U_1=U$ \cite{zhang2018,VanderWiel2002,Galpin2006,Ferreira2011}. 
One might expect that intra-dot Coulomb energy would be larger than inter-dot 
one. However, our results are valid as long as the inter-dot Coulomb 
interaction is not much smaller than the intra-dot interactions. Specifically 
(as we discuss in Sec. \ref{current}), the interaction strength defines 
different transport regimes. Our results are thus qualitatively valid as long 
as $U_0$ and $U_1$ lie within the same regime.  

The leads are modeled as  free electron gases, and are characterized by their 
Fermi functions with temperatures $T_{L(R)}$, and chemical potentials 
$\mu_{L(R)}=\mu_{av}\pm \Delta\mu/2$. 
Without loss of generality we consider different temperatures such that 
$T_R>T_L$, and define the voltage as $\Delta\mu=\mu_L-\mu_R$, 
which can be either positive or negative.

\subsection{Method}
\subsubsection{Rate Equations}
When considering the Hamiltonian of Eq. \ref{eq:Main_Hamiltonian} one may 
conceptually divide it into three parts: the DQD (system), the leads 
(reservoirs), and the tunneling between them. There are various ways to treat 
such systems, each has its advantages. 
In this work we use rate equations. Using rate-equation formalism, one finds 
the population of the system (diagonal elements of the density matrix) by 
solving a system of linear first-order differential equations. The standard 
procedure is to consider the steady state solution upon which the  problem 
reduces to an algebraic system of linear equations whose solution, i.e. the 
kernel of the rate matrix, is the population of the DQD. The rate-equations 
are valid in the weak couping limit which requires 
$\Gamma_{xij\sigma}\ll k_BT$\cite{Purkayastha2015,PhysRevB.82.235307}, where 
$\Gamma_{xij\sigma}(\omega) = \sum_k V_{kx i\si} V_{kx j\sigma}^* 
A_{kx\sigma}(\omega)$, and  $A_{kx\sigma}(\omega)$ is the spectral function.  
While generally $\Gamma_{xij\sigma}(\omega)$ is a function of energy ($\om$) 
\cite{PhysRevB.82.235307,MeirWingreenFor}, here we employ the wide band 
approximation, and assume constant density of states in the reservoirs within 
the relevant region of the spectrum, thus 
$\Gamma_{xij\sigma}(\omega)\cong\Gamma_{xij\sigma}$.

This approach is closely related to the master equations approach, and in fact 
they become the same at steady state for non-degenerate systems 
\cite{Schaller2014,breuer2002theory}. However, rate equations can also be 
thought of as a consequence of time-dependent perturbation theory in first 
order (Fermi golden rule) \cite{bruus2004many}, as a consequence of solving a 
many body \dinger equation \cite{Gurvitz1996}, or as limiting case of 
Non-Equlibrium Green-Function approach when the width of the spectral function  
vanishes. The main advantage of the use of rate equations is that with relative 
ease one can model quite complex systems, which would require much more effort 
in other approaches, yet one is able to observe many of the interesting 
phenomena that such systems demonstrate. Additionally, they allow treatment of 
systems with competing energy scales with relative ease. 

Within the rate-equation formalism we treat the DQD by diagonalizing 
$\Ham_{DQD}$ (diagonal in position basis in our case), while the leads are 
treated as electronic reservoirs, and are characterized by their Fermi 
distributions. Electron transfer between the DQD and either lead is treated 
only within the rate matrix $W$ whose  elements are the rates for transition 
between eigenstates of the reduced (sub-)system which is the DQD. The rate 
equations read \cite{breuer2002theory,bruus2004many,CartesanTim2008}:
\begin{equation}\label{eq:rate_eq}
\D{p_m}{t}=\sum_{\substack{n \\ n\neq m}}W_{n\to m}p_n-p_m
\sum_{\substack{n \\ n\neq m}}W_{m\to n}\;\;\Leftrightarrow\;\;
\dot{\mathbf{p}}=W\mathbf{p},
\end{equation}
where $W_{n\to m}=W_{mn}$ (for off-diagonal matrix elements) is the rate of 
transition from the many body Fock state $\ket[\big]{n}$ to $\ket[\big]{m}$, 
and $p_m$ - the probability that the system will be in the many-body state 
$\ket[\big]{m}$ of the DQD. The rate associated with adding an electron to 
the system is:
\begin{align}
\label{eq:additionRate}
		W_{n\to m} &=
        		 \sum_{x=L,R}	f(\epsilon_m-\epsilon_n-\mu_x) 	\times	\\
                 &\times \sum_{ \substack{\sigma \\ i,j=A,B}}	
                 \Gamma_{xij\sigma}		\transition{m}{d_{i\sigma}^\dagger}{n}	
                 \transition{n}{d_{j\sigma}}{m},	 \nonumber
\end{align}
while for removal of an electron the associated rate is:
\begin{align}
\label{eq:removalRate}
W_{n\to m} &=	 \sum_{x=L,R}	\Big(1-	f(\epsilon_n-\epsilon_m-\mu_x) 	
\Big)\times\\
& \times \sum_{ \substack{\sigma \\ i,j=A,B}}	\Gamma_{xij\sigma}		
\transition{m}{d_{j\sigma}}{n}	\transition{n}{d_{i\sigma}^\dagger}{m},	 
\nonumber
\end{align}
where $\epsilon_m$ is the energy of the $m$-th state of the DQD,  
$f(\epsilon-\mu_x)$ stands for the Fermi function of the $x$-th lead, and 
therefore represents the probability of the lead to have an electron with the 
energy of the desired quasi-level and $\Gamma_{xij\sigma}$ (defined above) is 
the coupling of the DQD to the leads. 
The overlaps of the form $\transition[\big]{m}{d_{j\sigma}}{n}$ dictate 
whether the transition is allowed in the sequential tunneling regime, where 
allowed transitions move the system between states which differ by one electron. 
Eq. \ref{eq:rate_eq} also implies that the diagonal elements of the rate matrix 
read:
\begin{equation}
	W_{mm} =  - \sum_{\substack{n \\ n\neq m}}W_{m\to n} =  
	- \sum_{\substack{n \\ n\neq m}}W_{nm},
\end{equation}
such that the sum of elements along each column vanishes.
Steady state populations are found by solving $W\mathbf{p}=0$, and imposing 
probability conservation ($\sum_m p_m=1$).

\subsubsection{Current}
By defining the total particle number in the DQD as 
$\hat{n}=\sum_{i\si}\hat{n}_{i\si}$, we may find the particle current 
by writing:
\begin{equation}
\begin{split}
	&		\D{\expect[\big]{\hat{n}}}{t}=\D{}{t}\left(\sum_m n_m p_m \right) = 
	\sum_m n_m \dot{p}_m = \sum_m n_m \left(Wp\right)_m  =		\\
	&		= \underbrace{\sum_m n_m \left(W^Lp\right)_m}_{\equiv J_L} + 
	\underbrace{\sum_m n_m \left(W^Rp\right)_m}_{\equiv J_R}  = 0,
\end{split} 
\end{equation}
where $W=W^L +W^R$, and $W^{L(R)}$ is the matrix which corresponds to rates 
involving only the left (right) lead, and $n_m=\transition[\big]{m}{\hat{n}}{m}$ 
(since $[\hat{n},\Ham_{DQD}]=0$ the value of $n_m$ is time independent as long 
as the leads aren't considered).
At steady state, the latter is trivially $J_L+J_R=0$, as the current flowing 
from the DQD to one lead, is exactly canceled by the current flowing into the 
DQD from the other. However, if we only consider the current between the DQD and 
one of the leads the expression does not vanish and we can express the actual 
electric current flowing through the system by:
\beq
J=J^L=\frac{e}{\hbar}\sum_m n_m \left(W^Lp\right)_m .
\eeq
The power harvested by the DQD can simply be found by:
\begin{equation}\label{eq:power_def}
P_{out}=-J\Delta\mu.
\end{equation}

In this work we express all energy scales in units of temperature. For quantum 
dots, typical operating temperatures range from milikelvins to room temperature, 
and our results consider temperatures of up to $1$K, as this is the range of 
temperatures in which relevant experimental work \cite{Svilans2016} was 
conducted.%
\section{Non-Monotononic Current-temperature bias relations} \label{current}
We begin the results section of the paper by describing the non-trivial  
dependence of the current on the temperature bias. Similar to negative 
differential conductance (NDC), upon
considering an interacting system with distinguishable transport 
channels  ($\Delta\epsilon>T$), one may observe negative thermal 
response (NTR), seen as a non-monotonic dependence of $J$ on 
thermal bias ($\Delta T$). 
However, the mechanisms responsible for the two phenomena are 
not the same. As discussed in 
\cite{Yoni_NDC,PhysRevB.76.035432}, in order to observe NDC one 
must consider unequal couplings between DQD and leads adhering 
to certain conditions. Even though NTR may be observed in the 
same  regime, it may also be observed with equal couplings, as 
shown in recent theoretical studies 
\cite{Sierra2014,Sierra2016}, and experimental work 
\cite{Svilans2016} (which includes also a rate equations model), and is related to a dynamical channel blockade in quantum dots \cite{belzig2005}.

\begin{figure*}
	\centering
	\begin{subfigure}{0.2\textwidth}
		\includegraphics[width=\textwidth]{{{fig_2a_illustrations}}}
		\caption{\label{fig:Basic_NTR_muL30_illus}}
	\end{subfigure}
    	\hspace{0.2cm}
	\begin{subfigure}{0.75\textwidth}
		\includegraphics[width=\textwidth]{{{fig_2b_Separate_Currents}}}
	\caption{\label{fig:Basic_NTR_muL30_Sep_curr}}
	\end{subfigure}
    \begin{subfigure}{0.42\textwidth}
		\includegraphics[width=\textwidth]{{{fig_2c_Alternative_prob}}}
		\caption{\label{fig:Basic_NTR_muL30_prob}}
	\end{subfigure}
	\hspace{0.2cm}
	\begin{subfigure}{0.47\textwidth}
		\includegraphics[width=\textwidth]{{{fig_2d_contour}}}
		\caption{\label{fig:Basic_NTR_muL30_contour}}
	\end{subfigure}
    \caption[]{
	\label{fig:Basic_NTR_muL30}%
	\subref{fig:Basic_NTR_muL30_illus}  Illustration of transport channels 
    alignments with the two chemical potentials $\mu_{L,R}$, where purple 
    arrows indicate the direction of 
    current through each channel. \subref{fig:Basic_NTR_muL30_Sep_curr} Current 
    through each dot ($J_{A,B}$) and total current ($J=J_A+J_B$) vs. the 
    temperature difference
    $\Delta T$ for two values of the interaction strengths: $U=100T,300T$. 
    \subref{fig:Basic_NTR_muL30_prob} Current 
    through the DQD ($J$), vs. probability of the DQD to contain two 
    particles ($p_2$), for various values of $U$. 
    \subref{fig:Basic_NTR_muL30_contour} 
    $J$ vs. 
    $\Delta T$ and $U$. The gray and blue vertical lines indicate values of $U$ 
    for which cross-sections were plotted in 
    \subref{fig:Basic_NTR_muL30_Sep_curr}, while the black curve
	 indicates the path along which the current has a local maxima 
	 (when plotted only against $\Delta T$). 
	Parameter values: $\Gamma=T/10,\; \epsilon_A=10T,\; \epsilon_B=40T,\; 
    \mu_L=20T,\; \mu_R=30T$.
}
\end{figure*}

Below, we discuss one of several configurations in which NTR is predicted in 
the presence of interactions, ranging from $U\sim \Delta\epsilon$ to strong 
interactions, $U\gg |\epsilon_{A,B}|,|\mu_{L,R}|,T$ (we measure energies 
and chemical potentials the empty dot, which is defined as 
zero energy), and with equal couplings. 
Due to the variety of parameters in our model, there are of course other configurations 
which may lead to the described phenomena. To give better intuition regarding the effect and to improve physical understanding, we try to shed light on the minimal requirements needed to observe the effect at the end of this section, and in appendix \ref{app:minimal_model}.

Unlike previous works \cite{Svilans2016,Sierra2014,Sierra2016}, we are 
motivated by potential applications regarding energy harvesting, and therefore 
consider also finite voltage ($\Delta\mu=\mu_L-\mu_R\neq0$).
To this end, $J$ is plotted as a function of $\Delta T$ such that 
$T_L=T,\;\;T_R=T+\Delta T$. The thermal bias needed in order to observe the 
phenomena in the high temperature limit ($k_BT\gg\Gamma$), is rather large and 
challenging from an experimental point of view as discussed by 
\cite{Svilans2016}, and our analysis assumes the maximal thermal bias as 
$\max(\Delta T)=40T$ as reported in their work, though we predict the phenomena 
to be evident even for smaller thermal bias.

Fig. \ref{fig:Basic_NTR_muL30_illus} illustrates the discussed configuration. 
It shows the alignment of leads with respect to accessible transport channels 
pertaining to single-particle states ($\epsilon_{A,B}$ in the figure), and to 
two-particle states ($\epsilon_{A,B}+U$ in figure), while  purple arrows 
indicate the direction of current flowing through each channel upon  sufficient 
heating of the right lead. Transport via single particle channels ($\epsilon_{A,B}$) occurs as an electron enters a previously empty system and then tunnels into the other lead. Transport via two-particle channels occurs as an electron tunnels into a system previously occupied with a single electron, thus requiring more energy ($\epsilon_{A,B}+U$) to tunnel into the system. Currents flowing through each dot ($J_{A,B}$) as 
well as total current ($J=J_A+J_B$) are plotted vs. thermal bias for two 
interaction strengths in Fig. \ref{fig:Basic_NTR_muL30_Sep_curr}.

 As expected, 
for $\Delta T=0$ no current flows as there is no available channel within the 
Fermi window. A small thermal bias ($\Delta T\sim5T$ in the figure) induces 
thermal current  only through single-particle channels. However, current through 
the lower channel (illustrated as $\epsilon_A$) prevails over current through 
the higher one (labeled as $\epsilon_B$) for the following reason: since 
$f_{L,R}(\epsilon_B)\leq \half \leq f_{L,R}(\epsilon_A)$ the probability of 
the system to be in Fock states with a single particle in dot $A$ is 
necessarily larger than  those with a particle in dot $B$ as insertion and 
extraction rates are proportional to $f$ and $1-f$. Due to the 
interaction, these channels cannot support current simultaneously, and current 
through $\epsilon_A$ prevails. Note that this dominance is maintained even 
though we chose $\mu_R$ such that it is closer to $\epsilon_B$, i.e. 
$|\mu_R-\epsilon_B|<|\mu_R-\epsilon_A|$, and thus one would expect  naively the 
effect of $\Delta T$ on current flowing through $\epsilon_B$ to be greater. 

For larger thermal bias, depending on the magnitude of $U$, current may also 
flow through higher channels  labeled as $\epsilon_{A,B}+U$ in Fig. 
\ref{fig:Basic_NTR_muL30_illus}. Current flowing through $\epsilon_B+U$ flows 
in the same direction as current through $\epsilon_B$  (since both 
$\epsilon_B,\epsilon_B+U>\mu_{L,R}$) and the two contributions add up. This is 
in contrast to current through $\epsilon_A$ and $\epsilon_A+U$ which flow in 
opposite directions (since $\epsilon_A<\mu_{L,R}<\epsilon_A+U$), which 
manifests in Fig. \ref{fig:Basic_NTR_muL30_Sep_curr} as a local maxima in the 
gray dashed curve. The suppression  of $J_A$ together with the enhanced $J_B$ 
are responsible for the local maxima in $J$ in the solid gray curve. 
Since the position of the maxima in $J$ vs. $\Delta T$ increases monotonically 
with $U$, the maxima in the blue line ($U=300T$) is shifted to a value of 
$\Delta T$ beyond the domain depicted in the figure.

The Coulomb interaction determines the occupation of different levels. It is 
thus interesting to look at the correspondence between changes in current 
line-shape and the  probability of 
the system to be in two-particle states. In Fig. 
\ref{fig:Basic_NTR_muL30_prob} $J$ is plotted vs. $p_2$ for various 
values of $U$, where $p_2$ is defined as sum of probabilities for being in  any 
of the two-particle states.
 The curves were obtained by finding both $J$ 
and $p_2$ for $\Delta T$'s within the domain $\Delta T\in (0,40T)$, and the 
curves terminate at the maximal allowed thermal bias, i.e. $\Delta T=40T$ 
(intuitively, $p_2$ increases monotonically with $\Delta T$). 
As shown, the
current increases as a function of $p_2$, up until a $U$ dependent threshold,  
at which it drops sharply with the onset of the NTR. Quite surprisingly, 
Fig.~\ref{fig:Basic_NTR_muL30_prob} shows a huge sensitivity of the current to 
the occupations; a seemingly insignificant probability of $1\%$ of the system 
to be in a two particle state drastically alters transport, and is enough 
to cause NTR.

 Fig. \ref{fig:Basic_NTR_muL30_contour} shows the broader picture via a contour 
 map of $J$ vs. $\Delta T$ and $U$. The black curve
shows the path along which the current has a local maxima for the specified 
$\Delta\epsilon,\mu_{L,R}$. The curve, which has been found numerically,
illustrates approximately the minimal thermal bias required for NTR to be 
observed for a given value of $U$.

As expressed before, the specific setting under discussion is just one of many to cause non-monotonic current, and as shown in the appendix, it is predicted even for simpler systems. We wish to summarize here the basic requirements needed for non-monotonic dependence of current on thermal bias:
\begin{description}
	\item[Alignment of Channels] The different behavior at finite thermal bias stems from competing current flow through different channels. Specifically, heating the electrode must create more electrons at an energy of one channel, while creating more holes at an energy of another. Thus, the chemical potential of the hot lead ($\mu_R$) must be higher than the energy of some accessible transport channels ($\mu_R>\epsilon_A$ in Fig. \ref{fig:Basic_NTR_muL30_illus}), and lower than others ($\mu_R<\epsilon_B,\epsilon_{A,B}+U$ in Fig. \ref{fig:Basic_NTR_muL30_illus}).
	%
	%
	\item[Competition] Competition between currents flowing in opposite directions via different transport channels can contribute to stronger non-monotonicity (though, as shown in the appendix, this is not a necessity). It happens if some channels are below both chemical potentials ($\mu_{L,R}>\epsilon_A$ in Fig. \ref{fig:Basic_NTR_muL30_illus}), while others are above both ($\mu_{L,R}<\epsilon_B,\epsilon_{A,B}+U$ in Fig. \ref{fig:Basic_NTR_muL30_illus}).
\end{description}


\section{Energy Harvesting}\label{harvesting}
Thermo-electric energy harvesting requires driving a current against a load, 
i.e. driving an electric current opposite to an external voltage ($\Delta\mu$) 
by means of a thermal bias ($\Delta T$) \cite{Sothmann2015}. 
Harvesting systems can be classified as n-type  if electric current flows from
 the hot lead to the cold one (same direction as heat flow), or as p-type if
  electric current flows from the cold lead to the hot one (opposite to heat 
  flow)\cite{Sothmann2015}, 
and we discuss the required settings for each regime, 
limiting the discussion to equal coupling and strong interactions. 
We then consider also the case of weaker interaction strength 
($U\sim\Delta \epsilon$).

\subsection{Monotonic Harvesting, N-Type}

Let us start with the case of very large (essentially infinite) $U$. 
For $\Delta\mu=\mu_L-\mu_R>0$ energy harvesting requires $J<0$ which occurs 
(approximately) when $f_R(\epsilon_A)> f_L(\epsilon_A)$. 
This is explained by the following: Trivially,  the direction of current 
through dot $A$ ($J_A$) depends solely on the above comparison. Similarly, the 
direction of current through $\epsilon_B$ is dictated by comparing 
$f_R(\epsilon_B)$ and $f_L(\epsilon_B)$. That being said,
for $\epsilon_B>\mu_L$ the current through $\epsilon_A$ prevails over current 
flowing through $\epsilon_B$, as, due to the strong interaction, 
the larger occupation of dot $A$  prohibits occupation and current flow through 
$\epsilon_B$. If, on the other hand, $\mu_L$ is assumed to be large enough such 
that occupations of the two dots are comparable, the currents through the two 
dots are similar too. Thus, in order to pass current against voltage 
(with $\Delta\mu>0$ and strong interactions) we require: 
$f_R(\epsilon_{A(B)})> f_L(\epsilon_{A(B)})\Rightarrow f_R(\epsilon_A)> 
f_L(\epsilon_A)$, since, as explained, the condition regarding $\epsilon_B$ is 
obeyed trivially if the one for $\epsilon_A$ does. 

The last inequality yields an analytic condition for harvesting:
\begin{equation}
\label{eq:harvest_relation_n_type}
	\begin{split}
		& f_R(\epsilon_A)> f_L(\epsilon_A)  \Rightarrow 
		\frac{1}{e^{\frac{\epsilon_A-\mu_R}{T+\Delta T}}+1}  >  
		\frac{1}{e^{\frac{\epsilon_A-\mu_L}{T}}+1} \\
		& \Rightarrow  \Delta T>\Delta T_{harvest} =
		\frac{T(\mu_L-\mu_R)}{\epsilon_A-\mu_L}  =  
		\frac{2T\Delta\mu}{2\epsilon_A-\mu_{av}-\Delta\mu}.
	\end{split}
\end{equation}
The above condition also contains another constraint, stating that for 
harvesting we must require $\epsilon_A>\mu_L$ as we consider $\Delta T>0$ and 
$\Delta\mu=\mu_L-\mu_R>0$. If this additional constraint is  not obeyed, no 
amount of heating will cause current  flow against the voltage (no harvesting) 
in the n-type regime with strong interactions (i.e. $U$ is large enough to 
prohibit transport via two-particle channels).

\begin{figure*}
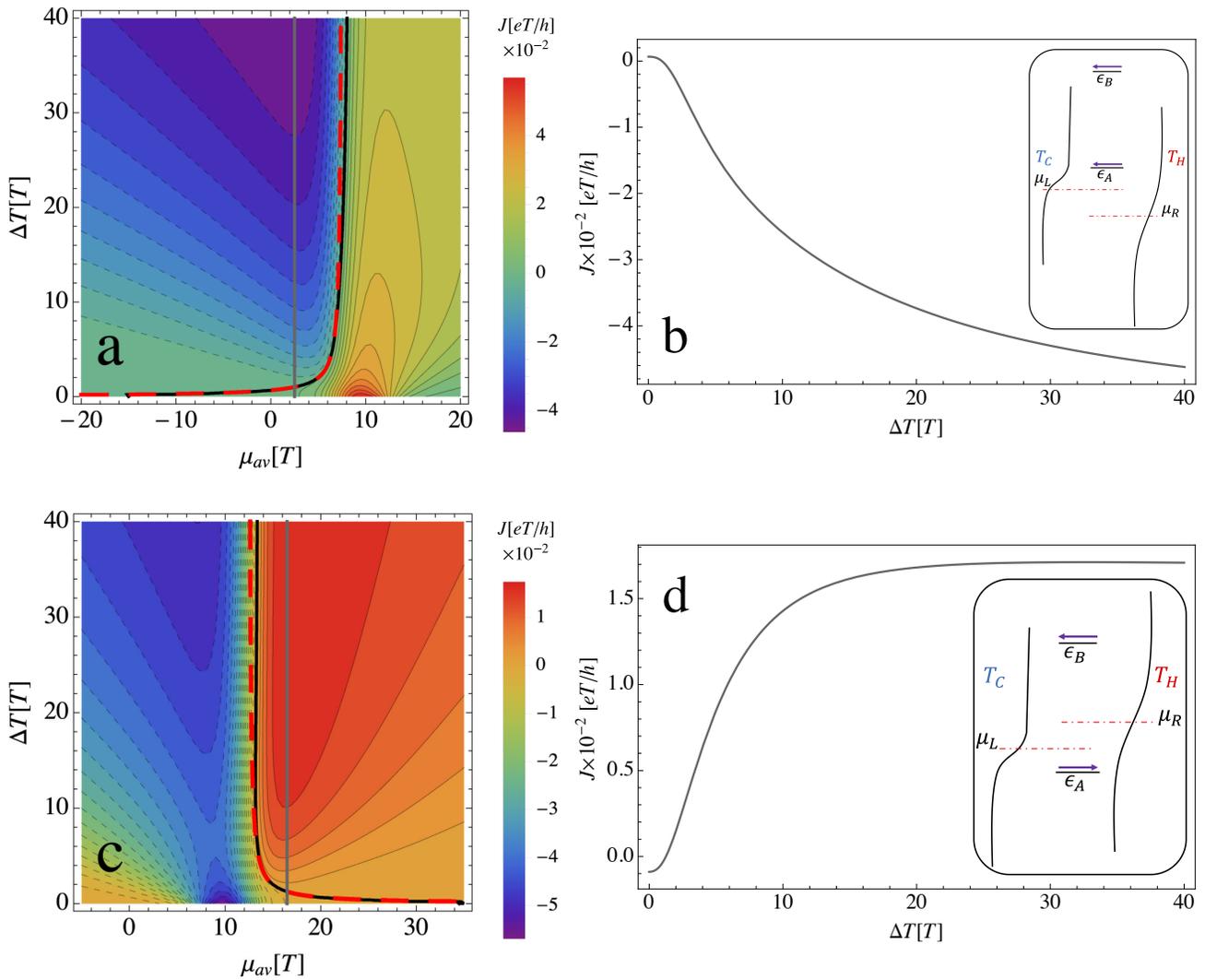

	\centering
	\begin{subfigure}{0.45\textwidth}
		\includegraphics[width=\textwidth]{{{fig_3a_N_type_contour}}}
		\caption{\label{fig:n_type_contour}}
	\end{subfigure}
	\begin{subfigure}{0.5\textwidth}
		\includegraphics[width=\textwidth]{{{fig_3b_N_type_current}}}
		\caption{\label{fig:n_type_current}}
	\end{subfigure}
	\begin{subfigure}{0.45\textwidth}
		\includegraphics[width=\textwidth]{{{fig_3c_P_type_contour}}}
		\caption{\label{fig:p_type_contour}}
	\end{subfigure}
	\begin{subfigure}{0.5\textwidth}
		\includegraphics[width=\textwidth]{{{fig_3d_P_type_current}}}
		\caption{\label{fig:p_type_current}}
	\end{subfigure}
	\caption[]{
		\label{fig:n_type_harvester}%
		\subref{fig:n_type_contour} Contour plot of $J$ vs. $\mu_{av}$ and 
		$\Delta T$ for $\Delta\mu>0$. 
		The black curve indicates the contour line along which $J=0$, 
		the red curve plots Eq. \ref{eq:harvest_relation_n_type} and the gray 
		vertical line ($\mu_{av}=2.5T$) indicates the path along which the cross 
		section in \subref{fig:n_type_current} is taken.
		\subref{fig:p_type_contour} Contour plot of $J$ vs. $\mu_{av}$ and 
		$\Delta T$ for $\Delta\mu<0$. The black curve indicates the contour 
		line along which $J=0$, the red curve plots 
		Eq. \ref{eq:harvest_relation_n_type} and the gray vertical 
		line ($\mu_{av}=16.5T$) indicates the path along which the cross 
		section in \subref{fig:p_type_current} is taken.
		Parameter values:
		$U=1000T,\; \Gamma=T/10,\; \epsilon_A=10T,\; \epsilon_B=40T,\; 
		|\Delta\mu|=5T$.
		}
\end{figure*}

Fig. \ref{fig:n_type_contour} shows a contour plot of $J$ vs. $\mu_{av}$ and 
$\Delta T$ for fixed $\Delta\mu$. Dashed contour lines indicate the domain in 
which current is negative (the DQD acts as a harvester, since $\Delta\mu>0$), 
while solid contour lines indicate the domain of positive current, and the two 
are separated by the thick black curve indicating the curve along which current 
vanishes ($J=0$).\hspace{0.1cm} The thick red  curve shows the analytic 
condition derived above (Eq. \ref{eq:harvest_relation_n_type}). As can be seen, 
the above condition approximates wery well the black curve. The deviation at 
large $\Delta T$ stems from  the fact that our derivation does not take in to 
account the  less significant current flowing through $\epsilon_B$.

Fig. \ref{fig:n_type_current} shows a cross section of the contour plot in 
Fig. \ref{fig:n_type_contour} taken at $\mu_{av}=2.5T$ as indicated by the 
vertical gray line in the contour map. In this regime the current is monotonic 
with respect to $\Delta T$, in agreement with the discussion in the previous 
section,  as there is no competition between currents flowing in opposite 
directions. This is shown in the  inset, which illustrates the alignment of the 
leads chemical potentials with the transport channels, with  arrows indicating 
the direction of current flow through each channel upon heating of the right 
lead.

\subsection{Monotonic Harvesting, P-Type}

The DQD may also act as a p-type harvester and drive electric current from the 
cold lead to the hot one against a voltage, with the previous inequality merely 
changing direction, i.e. $f_L(\epsilon_A)>f_R(\epsilon_A)$. Therefore the 
relation in Eq. \ref{eq:harvest_relation_n_type} still determines the cross-over 
between harvesting and investing energy, with the only change being that now we 
consider $\Delta\mu<0$, and thus the trivial condition switches to 
$\epsilon_A<\mu_L$.

Fig. \ref{fig:p_type_contour} shows a contour plot of $J$ vs. $\mu_{av}$ and 
$\Delta T$, with dashed contours indicating negative current, and solid contours 
indicating positive current (here harvesting requires $J>0$ as we consider 
$\Delta\mu<0$). The two domains are separated by the thick black curve 
indicating the curve along which current vanishes ($J=0$), while the thick red 
one shows the analytic approximation derived above (Eq. 
\ref{eq:harvest_relation_n_type}). As before, the red curve slightly deviates 
from the black one for large $\Delta T$, stemming from neglecting the 
contribution to current through $\epsilon_B$. 
 
 Fig. \ref{fig:p_type_current} shows a cross section of the contour plot in Fig. 
 \ref{fig:p_type_contour} taken at $\mu_{av}=16.5T$ as indicated by the vertical 
 gray line in the contour map. The inset illustrates the alignment of the leads 
 chemical potentials with the transport channels, and the arrows indicate the 
 direction of current flow through each channel upon heating of the right lead.  
 Here too, the current is monotonic with respect to $\Delta T$. Since the 
 voltages taken into account in figures \ref{fig:n_type_current} and 
 \ref{fig:p_type_current} are of the same magnitude, comparing current 
 magnitudes reveals that for the discussed configurations the n-type regime 
 allows for more power to be harvested.

\subsection{Non-Monotonic Harvesting}
\subsubsection{Interaction Facilitated Harvesting}
\begin{figure}[h]
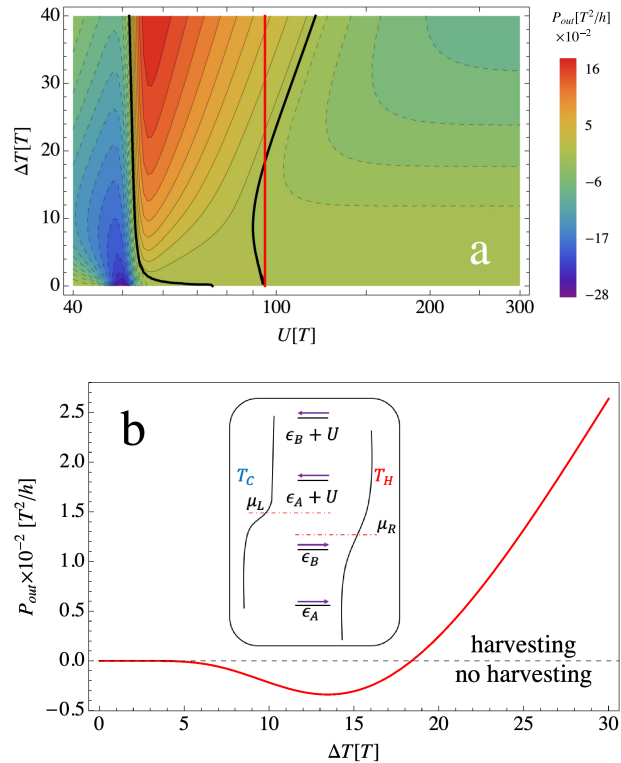

	\centering
		\begin{subfigure}{0.45\textwidth}
			\includegraphics[width=\textwidth]{{{fig_4a_N_type_contour}}}
			\caption{\label{fig:finite_U_N_type_contour}}
		\end{subfigure}
		\hspace{0.1cm}
		\begin{subfigure}{0.45\textwidth}
			\includegraphics[width=\textwidth]{{{fig_4b_N_type_current}}}
			\caption{\label{fig:finite_U_N_type_current}}
		\end{subfigure}
	\caption[]{
		\label{fig:finite_U_N_type}%
		\subref{fig:finite_U_N_type_contour} Contour plot of the output power $P_{out}$ vs. $U$ and 
		$\Delta T$. The black curves indicates the contour line along which 
		$P_{out}=0$, the red vertical line ($U=95T$) indicates the path along which 
		the cross section in \subref{fig:finite_U_N_type_current} is taken. Harvesting occurs for $P_{out}>0$ as indicated in the 
		cross section.
		Parameter values:
		$\Gamma=T/10,\; \epsilon_A=10T,\; \epsilon_B=40T,\; \Delta\mu=5T,\; 
		\mu_{av}=60T$.
	}
\end{figure}

Allowing for a small enough interaction such that two-particle states may get 
occupied and current may flow through two-particle channels has trivial as well 
as surprising effects regarding harvesting. Trivially, as current flows through 
more channels the dominance of flow through $\epsilon_A$ is reduced, thus the 
analytic approximation employed before becomes less relevant. Another effect is 
that, if the bare levels are submerged in Fermi sea (single-particle channels 
are inaccessible),  considering different strengths of interaction ($U$) may 
dictate whether the system harvests energy or not upon heating. This is easily 
seen in Fig. \ref{fig:finite_U_N_type_contour} which shows the power output, $P_{out}=I \Delta \mu$ vs. $U$ and 
$\Delta T$ for an n-type configuration ($\mu_L-\mu_R=5T$). As shown, the 
question whether or not the system harvests energy ($P>0$, solid 
contour lines) depends on $U$ and $\Delta T$. More notably, there are values of 
$U$ (for instance $U=95T$ marked by a red vertical line in Fig. 
\ref{fig:finite_U_N_type_contour}) for which the system's behavior changes 
non-monotonically with $\Delta T$ between harvesting and investing energy 
(Fig. \ref{fig:finite_U_N_type_current}). 

\subsubsection{Harvesting via Single-Particle Channels}

\begin{figure}[b]
	\vskip 0.5truecm
	\centering
			\includegraphics[width=0.65\textwidth]{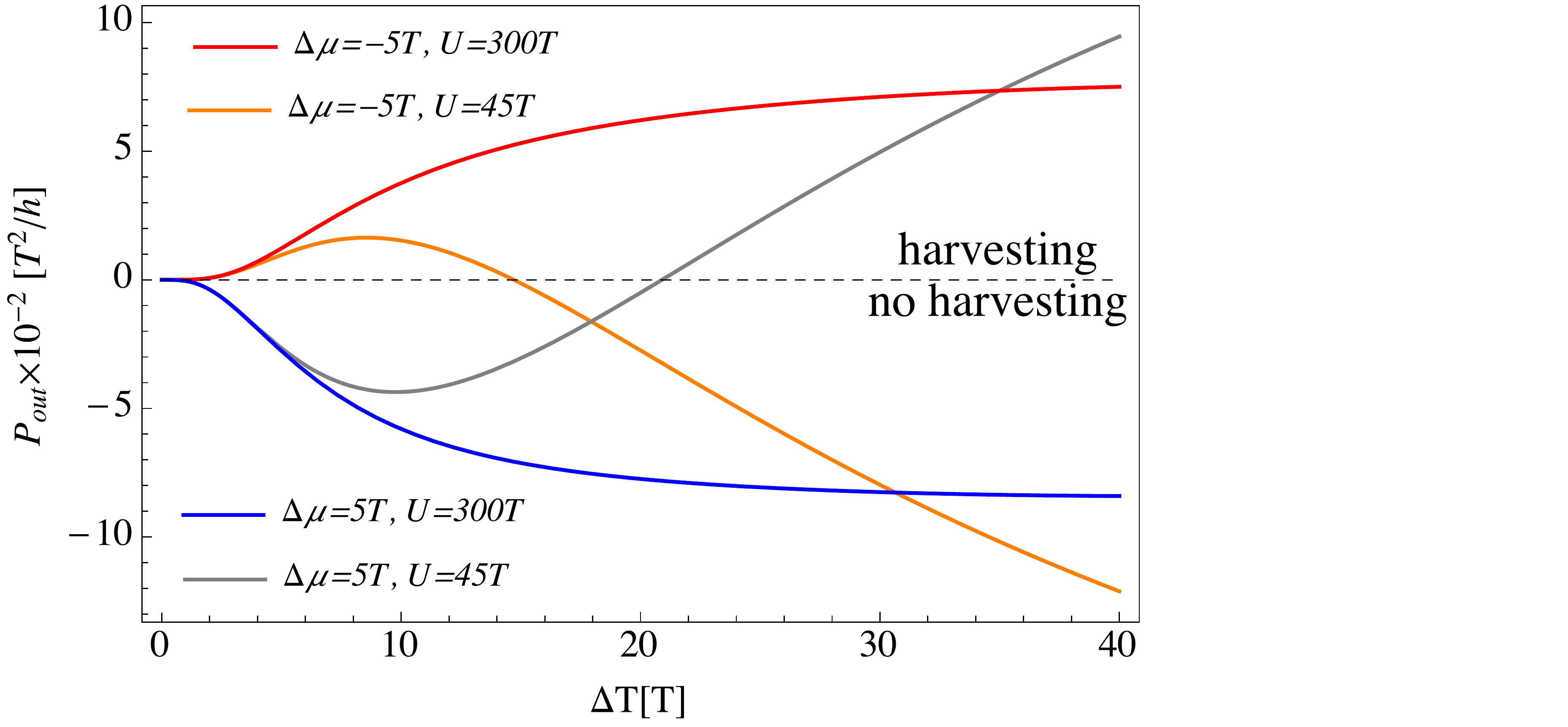}
	\caption[]{
		\label{fig:finite_U_power}%
		$P_{out}$ vs. $\Delta T$ for given values of $\Delta\mu$ and $U$. 
		$P_{out}>0$ means that the system harvests energy.
		Parameter values:
		$\Gamma=T/10,\; \epsilon_A=10T,\; \epsilon_B=40T,\; 
		|\Delta\mu|=5T,\; \mu_{av}=25T$.
	}
\end{figure}

When transport is dominated by current flowing through single-particle channels, 
interaction may either assist or  hinder harvesting depending on configuration. 
To understand why this is the case, one must note that in the n-type regime 
harvesting stems from current via channels which are aligned above the hot 
lead chemical potential ($\epsilon_i>\mu_R$), as illustrated in Fig. 
\ref{fig:n_type_current}. On the other hand, in the p-type regime, harvesting 
stems from current via channels which are beneath the hot lead chemical 
potential ($\epsilon_i<\mu_R$), as illustrated in Fig. \ref{fig:p_type_current}. 
This is shown in Fig. \ref{fig:finite_U_power}, where the output power $P_{out}$ 
(Eq. \ref{eq:power_def}) is plotted as a function of $\Delta T$ for opposite 
voltages (n-type and p-type regimes), and for two interactions strengths, 
$U=45T,300T$. The relevant configuration is the same as the one illustrated in 
Fig. \ref{fig:Basic_NTR_muL30_illus}. As seen, in the n-type regime 
($\Delta\mu>0$) the two-particle channels assist harvesting and in fact if 
$U$ is too large harvesting is impossible in this configuration as the 
two-particle channels are blocked. On the other hand, in the p-type regime 
($\Delta\mu<0$), the interaction hinders harvesting, and therefore if the 
interaction is small enough to allow two-particle channels to affect transport, 
harvesting is possible only for a limited ($U$ dependent) range of thermal bias.

\subsubsection{Maximal power and optimal thermal bias}
\begin{figure}
	\centering
	\includegraphics[width=0.5\textwidth]{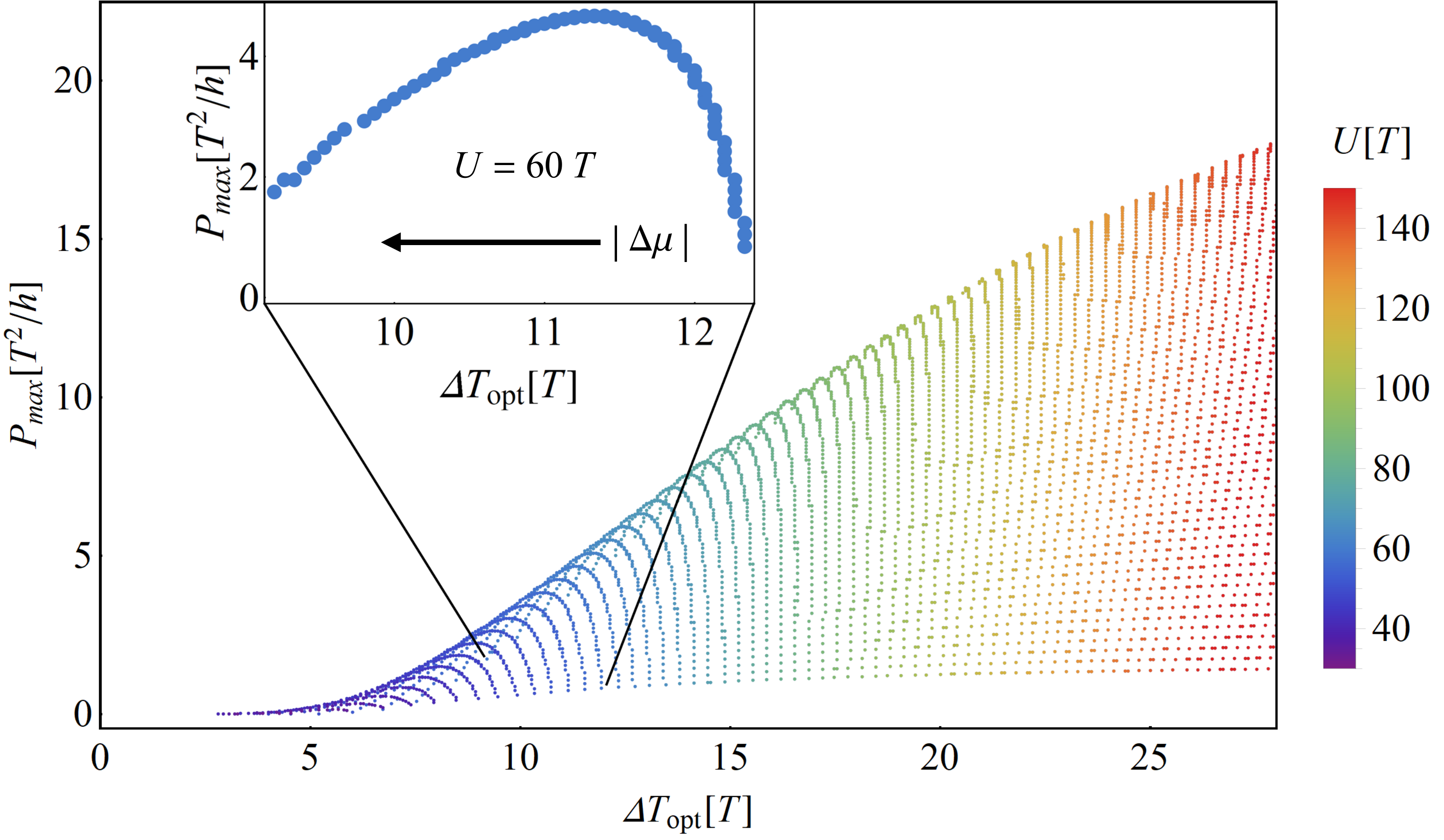}
	\caption[]{
		\label{Fig_Pmax}%
Maximal power output $P_{max}$ plotted against optimal temperature difference 
$\Delta T_{opt}$, calculated for different voltage biases and interaction 
strengths (see text for numerical paramters). The key observation is that, as 
opposed to non-interacting system, there is an optimal bias voltage and 
temperature difference for which power output is maximal. Inset: selected data 
set for $U=60T$, showing the power-temperature non-monotonic curve.  
	}
\end{figure}

As implied from the previous discussion, under given conditions (negative 
bias, so called p-type regime, and intermediate interaction strength) the power 
output displays non-monotonic behavior as a function of the thermal bias. This 
needs to be accounted for, if one wishes to optimize the performance of the 
thermoelectric device as an energy harvester. To explore this situation, we have 
performed the following computation. A set of DQD systems, all with the same 
energies, average chemical potential and couplings, but with different 
interaction strengths ($30T<U<150$)  were considered 
($\epsilon_A=10T,\epsilon_B=40T, \mu_{av}=25T,\Gamma=T/10$). The  bias voltage 
was scanned in the range $-20T<\Delta \mu<-T$. For each system (value of $U$)
 and each voltage bias ($\Delta \mu$), the temperature bias $\Delta T$ for 
 which one obtains maximal power was found. Each point in Fig.~\ref{Fig_Pmax} 
 depicts that maximal power as a function of the optimal thermal bias for the 
 different voltage biases, and a given value of $U$.
The different curves correspond to different interaction strengths, according 
to the color coding. 

The first thing one notices is that the larger the 
interaction strength, more power can be extracted at optimal conditions. This 
is easy to understand; in the discussed setting, the harvested current passes
 through the lower transport channel ($\epsilon_A$). The current that flows through two particle channels
(at energies $\epsilon_{A(B)}+U$), in the direction of the voltage bias (suppressing harvesting), is diminished by the increasing interaction. 

Second, for each value of $U$ there is an optimal thermal bias ($\Delta T_{opt}$) which maximizes 
the harvested power. The exact $\Delta T$ for which power is maximal has a non-trivial dependence on $U$, and in fact, as discussed in the appendix, even for a non-interacting system the maximum of harvested power has no simple analytic form.

Third, one can observe a specific structure of the data; for each value of $U$ there is a 
well-defined "trajectory" when plotted against $\Delta T_{opt}$. The inset shows 
one example of such a trajectory 
(taken at $U=60T$), in the $\{P_{max}, \Delta T_{opt}\}$ plane. The arrow indicates 
the direction of increasing negative voltage, from $\Delta\mu=-1T$ to $\Delta\mu=-20T$. Maximal 
power appears at $\Delta \mu \sim -11T$. In general, the optimal power must 
depend on the applied voltage. This is in contrast with the strong-$U$ case (as in Fig. \ref{fig:n_type_harvester}), where one expects a monotonic dependence on voltage. For the (intermediate) interacting case, on the 
other hand, the non-monotonicity in the current implies a more subtle relation 
between optimal power, optimal thermal bias and voltage bias, depicted in 
Fig.~\ref{Fig_Pmax}. 
This point should be considered for optimal design of an experimental apparatus.

\section{Summary and Conclusions}\label{summary}

To summarize, in this work we studied the thermoelectric current and energy 
harvesting in an interacting double-quantum-dot system, connected to reservoirs 
held at different chemical potentials and temperatures. Using a rate-equation 
approach, the current was evaluated for different energetic configurations of 
the DQD. We discuss in details the current-temperature gradient relations 
(the thermoelectric analog to current-voltage relations) and the conditions 
under which energy (from the temperature gradient) can be harvested and 
converted to useful electrical power. 

Specifically, our main results are: (i) In the presence of interactions, the 
current-thermal bias relations may exhibit non-monotonic behavior, in analogy 
to negative differential conductance in voltage-biased interacting quantum dot 
junctions. (ii) there is an extreme sensitivity of the current to the 
populations on the dot. Specifically, a tiny change in the quantum dot 
occupations can alter qualitatively the current-bias voltage dependence. (iii) 
Energy harvesting can be enhanced or hindered by Coulomb interactions, 
depending on the specific conditions of the DQD. (iv) In the presence of 
interactions, the non-monotonic current leads to non-trivial optimal 
configuration of the DQD. More specifically, one can say that it is not always the case that the larger the thermal bias 
the more energy can be harvested. 

These results provide insight into the nature of thermoelectric transport in 
quantum-dot junctions. Being well within current experimental capabilities, our 
predictions can be tested experimentally. Going beyond the weak-coupling limit, 
extending the system to more than two quantum dots, and allowing for direct 
coupling between the quantum dots (i.e. electron tunneling between the dots) 
are all directions we plan to address in the future. 

\acknowledgments
YM acknowledges support from ISF grant 292/15. YD acknowledges support from ISF grant 1360/17.

\appendix
\section{Minimal Models for Non-Monotononic Current} \label{app:minimal_model}
In this appendix we address the question of what are the minimal requirements for non-monotonic current. We demonstrate similar phenomena to those discussed in the text can be expected for simpler systems, depending upon proper tuning of the system parameters. 

\subsection{Single Interacting QD}
\begin{figure}
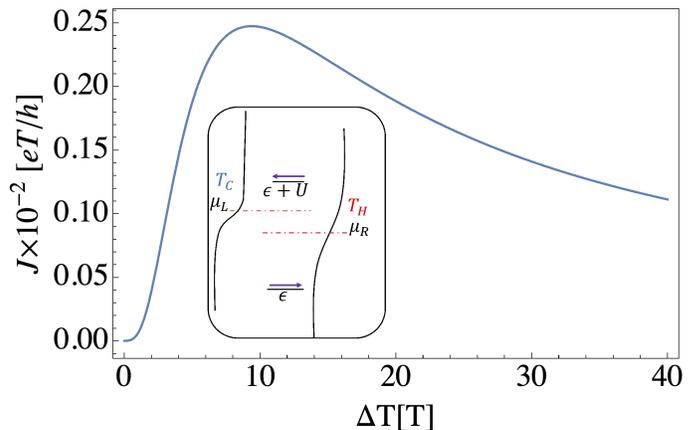

	\centering
	\includegraphics[width=0.5\textwidth]{{{fig7_singleQD}}}
	\caption[]{
		\label{fig:single_QD}%
		Current flowing through the QD vs. $\Delta T$. The inset illustrates the alignment of transport channels with respect to the chemical potential of the leads. Parameter values: 
		$U=30T,\; \Gamma=T/20,\; \epsilon=10T,\;  
		\mu_L=28T,\; \mu_R=22T$.
	}
\end{figure}

For a single, spinful QD (SQD), with Coulomb repulsion between spin-degenrate states, the Hamiltonian reduces to:
\begin{equation}
\label{eq:Single_QD_Ham}
\Ham_{QD}=\sum_{\si}\e d_\si^\dagger d_\si
+U \hat{n}_{\down}\hat{n}_{\up}  ~.
\end{equation}
Since the system still contains two transport channels (at energies $\e$ and $\e+U$), aligning the chemical potential of the leads as specified in Sec. \ref{current} such that $\e<\mu_R$ and $\e+U>\mu_R$, the currenct $J(\Delta T)$ versus finite  thermal bias ($\Delta T$) is non-monotonic. 
 This is shown in Fig. \ref{fig:single_QD}, which shows the current  flowing through the QD vs. $\Delta T$, with an illustration of the alignment of transport channels with respect to the chemical potentials of the leads in the inset.

\subsection{Non-Interacting DQD}
If one considers a non-interacting DQD the Hamiltonian reduces to:
\begin{equation}
\Ham_{DQD}=\sum_{ i=A,B}\e_i d_i^\dagger d_i
\end{equation}
The spin degree of freedom has been omitted from the above Hamiltonian as
 we wish to compare the following results with the results of  the single QD
  case presented before. Therefore we wish to consider a system which can contain up to two electrons simultaneously, at one of two energy levels.
  
  As long as we consider a non-interacting system we can treat it either by the rate-equation formalism utilized throughout this work, or by Landauer formula. Following the work of \onlinecite{PhysRevB.78.161406}, an expression for the transmission similar to theirs can be obtained, as well as an expression for current in terms of digamma functions.
  
\begin{figure}
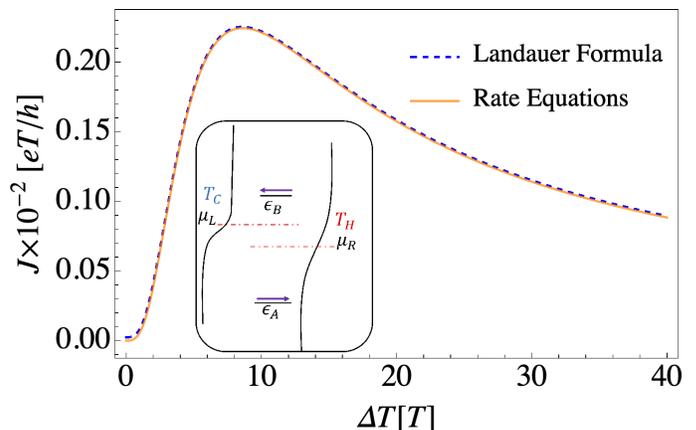

	\centering
	\includegraphics[width=0.5\textwidth]{{{fig8_non_interacting_DQD}}}
	\caption[]{
		\label{fig:non_interacting_DQD}%
		Current flowing through the DQD vs. $\Delta T$. The orange curve was obtained using rate equations, and the blue dashed curve was obtained by using Landauer formula.
		The inset illustrates the alignment of transport channles with respect to the chemical potential of the leads. Parameter values: 
		$\Gamma=T/20,\; \epsilon_A=10T,\; \epsilon_B=40T,\; 
		\mu_L=28T,\; \mu_R=22T$.
	}
\end{figure}

  Figure \ref{fig:non_interacting_DQD} shows the current flowing through the non-interacting DQD vs. 
  $\Delta T$. The figure shows results  using both rate equations and Landauer
   formula based expressions. As shown, since we consider weak coupling limit
   ($\Gamma\ll K_BT$), the results are equivalent. Unlike the results of 
   \onlinecite{PhysRevB.78.161406} we do not limit ourselves to linear response. 

\subsection{Roles of Interactions}
\begin{figure}[h]
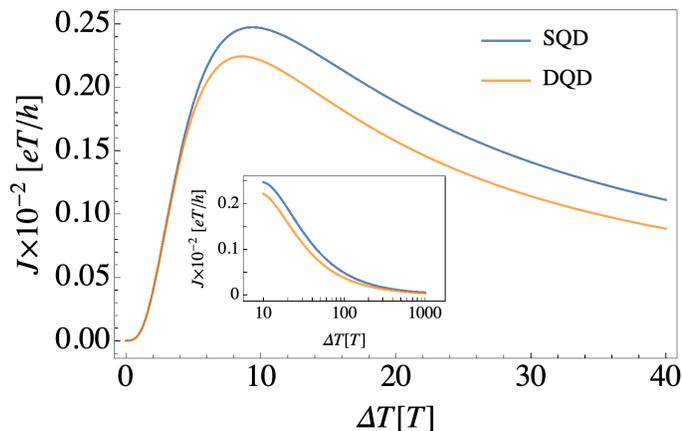

	\centering
	\includegraphics[width=0.5\textwidth]{{{fig9_Interaction_effects}}}
	\caption[]{
		\label{fig:Interaction_effects}%
		Current flowing through the system vs. $\Delta T$. The inset shows the same plot but for extremely high temperature difference. Parameter values: 
		$\Gamma=T/20,\; \epsilon_A=10T,\; \epsilon_B=40T,\; 
		\mu_L=28T,\; \mu_R=22T$.
	}
\end{figure}

To understand the effect of interaction on the system, in figure \ref{fig:Interaction_effects} we plot $J$ vs. $\Delta T$ for two systems. In blue we show the results of the single interacting QD (from Fig. \ref{fig:single_QD}), and in orange we show  the results of the non-interacting DQD  (from Fig. \ref{fig:non_interacting_DQD}). As shown, the line-shapes are similar, yet they differ even though both systems can occupy up to two electrons simultaneously, and even though both have two transport channels at the same energy. That being said, the systems differ if one considers the fact that the for the interacting system lower channel of the QD is spin-degenerate. This sort of behavior where degeneracy induces scaling  is not unique to our system and has been previously discussed \cite{PhysRevB.44.1646,eitan}. 

To conclude, in this example the interaction does nor affect the mechanisms described in a unique manner. However, as it affects the number transport channels and their degeneracy and  it enables some features which are nor predicted in their absence.  For instance, setups such as those realized in figures \ref{fig:finite_U_N_type_current} and \ref{fig:finite_U_power}, where the systems alternates between harvesting and investing energy cannot be realized in any of the simpler systems discussed in this appendix. This is due to the simple reason that even if we consider an infinitely large thermal bias, the current will not change its sign. Instead it will simply vanish altogether, as at infinite temperature the hot lead Fermi distribution is simply $f=1/2$ for all energies, which dictates zero current for a system with the same amount of channels which support current in opposite directions. This is seen in the inset of Fig. \ref{fig:Interaction_effects} in which thermal bias is allowed to be ridiculously high resulting in vanishing current.

A more fundamental consequence of interaction is shown in figure \ref{fig:Interaction_effects2}. Here, we again consider both the non-interacting DQD and the SQD, yet we align the chemical potentials such that the higher channel ($\epsilon_B$ or $\epsilon+U$ depending on the system) is within the Fermi window. Figure \ref{fig:Interaction_effects2} shows $J(\Delta T)$ for the described setup, and in this setup the fundamental difference between the systems is shown. For the DQD heating the right lead disturbs current flowing via $\epsilon_B$, but for larger $\Delta T$ it enables also current flow via $\epsilon_A$ which consequently results in non-monotonic line-shape. For the SQD, heating the right lead disturbs current flowing via $\epsilon +U$. However, enabling current flow via $\epsilon$ for larger $\Delta T$ reduces current flow via $\epsilon+U$ further, such that the overall flow cannot benefit from it.

\begin{figure}[h]
	\centering
	\includegraphics[width=0.5\textwidth]{{{fig10_Interaction_effects2}}}
	\caption[]{
		\label{fig:Interaction_effects2}%
		Current flowing through the system vs. $\Delta T$. The inset illustrates the alignment of transport channles with respect to the chemical potential of the leads. Parameter values: 
		$\Gamma=T/20,\; \epsilon_A=10T,\; \epsilon_B=40T,\; 
		\mu_L=28T,\; \mu_R=22T$.
	}
\end{figure}

Figure \ref{fig:Interaction_effects2} shows also two more general phenomena.  
When aligning the chemical potentials such that one of the channels is within he Fermi window, it is possible to; First, predict non-monotonic current without competition (i.e. current  through all channels flows in the same direction). Second, it is possible to predict a local minima for in the magitude of $J(\Delta T)$.  This can also be realized with an interacting system and not just for the non-interacting DQD.

\bibliographystyle{unsrt}
\bibliography{bibliography}

\begin{thebibliography}{10}

\bibitem{Mahan1996}
GD~Mahan and JO~Sofo.
\newblock The best thermoelectric.
\newblock {\em Proceedings of the National Academy of Sciences},
  93(15):7436--7439, 1996.

\bibitem{Dresselhaus1999}
MS~Dresselhaus, G~Dresselhaus, X~Sun, Z~Zhang, SB~Cronin, and T~Koga.
\newblock Low-dimensional thermoelectric materials.
\newblock {\em Physics of the Solid State}, 41(5):679--682, 1999.

\bibitem{Dresselhaus2007}
Mildred~S Dresselhaus, Gang Chen, Ming~Y Tang, RG~Yang, Hohyun Lee, DZ~Wang,
  ZF~Ren, J-P Fleurial, and Pawan Gogna.
\newblock New directions for low-dimensional thermoelectric materials.
\newblock {\em Advanced materials}, 19(8):1043--1053, 2007.

\bibitem{wang2014}
Xiaodong Wang and Zhiming~M Wang.
\newblock {\em Nanoscale thermoelectrics}, volume~16.
\newblock Springer, 2014.

\bibitem{cui2017}
Longji Cui, Ruijiao Miao, Chang Jiang, Edgar Meyhofer, and Pramod Reddy.
\newblock Perspective: Thermal and thermoelectric transport in molecular
  junctions.
\newblock {\em The Journal of Chemical Physics}, 146(9):092201, 2017.

\bibitem{benenti2017fundamental}
Giuliano Benenti, Giulio Casati, Keiji Saito, and Robert~S Whitney.
\newblock Fundamental aspects of steady-state conversion of heat to work at the
  nanoscale.
\newblock {\em Physics Reports}, 694:1--124, 2017.

\bibitem{Segal2005}
Dvira Segal.
\newblock Thermoelectric effect in molecular junctions: A tool for revealing
  transport mechanisms.
\newblock {\em Physical Review B}, 72(16):165426, 2005.

\bibitem{Li2016}
Yueqi Li, Limin Xiang, Julio~L Palma, Yoshihiro Asai, and Nongjian Tao.
\newblock Thermoelectric effect and its dependence on molecular length and
  sequence in single dna molecules.
\newblock {\em Nature communications}, 7:11294, 2016.

\bibitem{Korol2016}
Roman Korol, Michael Kilgour, and Dvira Segal.
\newblock Thermopower of molecular junctions: Tunneling to hopping crossover in
  dna.
\newblock {\em The Journal of Chemical Physics}, 145(22):224702, 2016.

\bibitem{Beenakker1992}
C.~W.~J. Beenakker and A.~A.~M. Staring.
\newblock Theory of the thermopower of a quantum dot.
\newblock {\em Phys. Rev. B}, 46:9667--9676, 1992.

\bibitem{Staring1993}
AAM Staring, LW~Molenkamp, BW~Alphenaar, H~Van~Houten, OJA Buyk, MAA Mabesoone,
  CWJ Beenakker, and CT~Foxon.
\newblock Coulomb-blockade oscillations in the thermopower of a quantum dot.
\newblock {\em EPL (Europhysics Letters)}, 22(1):57, 1993.

\bibitem{thierschmann2016thermoelectrics}
Holger Thierschmann, Rafael S{\'a}nchez, Bj{\"o}rn Sothmann, Hartmut Buhmann,
  and Laurens~W Molenkamp.
\newblock Thermoelectrics with coulomb coupled quantum dots.
\newblock {\em arXiv preprint arXiv:1603.08900}, 2016.

\bibitem{Erdman2017}
Paolo~Andrea Erdman, Francesco Mazza, Riccardo Bosisio, Giuliano Benenti,
  Rosario Fazio, and Fabio Taddei.
\newblock Thermoelectric properties of an interacting quantum dot based heat
  engine.
\newblock {\em Physical Review B}, 95(24):245432, 2017.

\bibitem{Whitney2014}
Robert~S. Whitney.
\newblock Most efficient quantum thermoelectric at finite power output.
\newblock {\em Phys. Rev. Lett.}, 112:130601, 2014.

\bibitem{Talbo2017}
Vincent Talbo, J{\'e}r{\^o}me Saint-Martin, Sylvie Retailleau, and Philippe
  Dollfus.
\newblock Non-linear effects and thermoelectric efficiency of quantum dot-based
  single-electron transistors.
\newblock {\em Scientific reports}, 7(1):14783, 2017.

\bibitem{Josefsson2018}
Martin Josefsson, Artis Svilans, Adam~M. Burke, Eric~A. Hoffmann, Sofia
  Fahlvik, Claes Thelander, Martin Leijnse, and Heiner Linke.
\newblock A quantum-dot heat engine operating close to the thermodynamic
  efficiency limits.
\newblock {\em Nature Nanotechnology}, 2018.

\bibitem{Kennes2013}
D.~M. Kennes, D.~Schuricht, and V.~Meden.
\newblock {Efficiency and power of a thermoelectric quantum dot device}.
\newblock {\em EPL (Europhysics Lett.}, 102(5):57003, 2013.

\bibitem{Sanchez2016}
David Sanchez and Rosa Lopez.
\newblock Nonlinear phenomena in quantum thermoelectrics and heat.
\newblock {\em Comptes Rendus Physique}, 17(10):1060--1071, 2016.

\bibitem{Thierschmann2015}
Holger Thierschmann, Rafael S{\'a}nchez, Bj{\"o}rn Sothmann, Fabian Arnold,
  Christian Heyn, Wolfgang Hansen, Hartmut Buhmann, and Laurens~W Molenkamp.
\newblock Three-terminal energy harvester with coupled quantum dots.
\newblock {\em Nature nanotechnology}, 10(10):854--858, 2015.

\bibitem{Svilans2016}
Artis Svilans, Adam~M. Burke, Sofia~Fahlvik Svensson, Martin Leijnse, and
  Heiner Linke.
\newblock {Nonlinear thermoelectric response due to energy-dependent transport
  properties of a quantum dot}.
\newblock {\em Phys. E Low-dimensional Syst. Nanostructures}, 82:34--38, 2016.

\bibitem{Svensson2013}
S~Fahlvik Svensson, Eric~A Hoffmann, Natthapon Nakpathomkun, P~M Wu, HQ~Xu,
  Henrik~A Nilsson, David S{\'a}nchez, Vyacheslavs Kashcheyevs, and Heiner
  Linke.
\newblock Nonlinear thermovoltage and thermocurrent in quantum dots.
\newblock {\em New Journal of Physics}, 15(10):105011, 2013.

\bibitem{Vicioso2018}
Alejandro Marcos-Vicioso, Carmen L\'opez-Jurado, Miguel Ruiz-Garcia, and Rafael
  S\'anchez.
\newblock Thermal rectification with interacting electronic channels:
  Exploiting degeneracy, quantum superpositions, and interference.
\newblock {\em Phys. Rev. B}, 98:035414, 2018.

\bibitem{Sothmann2015}
Bj{\"{o}}rn Sothmann, Rafael S{\'{a}}nchez, and Andrew~N Jordan.
\newblock {Thermoelectric energy harvesting with quantum dots}.
\newblock {\em Nanotechnology}, 26(3):032001, 2015.

\bibitem{Sierra2016}
Miguel~A. Sierra, M.~Saiz-Bret{\'{i}}n, F.~Dom{\'{i}}nguez-Adame, and David
  S{\'{a}}nchez.
\newblock {Interactions and thermoelectric effects in a parallel-coupled double
  quantum dot}.
\newblock {\em Phys. Rev. B}, 93(23):235452, 2016.

\bibitem{Gong2012}
Wei-Jiang Gong, Cui Jiang, Xiaoyan Sui, and An~Du.
\newblock Thermoelectric properties in a parallel double quantum dot structure
  modulated by the fano interferences.
\newblock {\em Journal of the Physical Society of Japan}, 81(10):104601, 2012.

\bibitem{Chi2011}
Feng Chi, Jun Zheng, Xiao-Dong Lu, and Kai-Cheng Zhang.
\newblock Thermoelectric effect in a serial two-quantum-dot.
\newblock {\em Physics Letters A}, 375(10):1352--1356, 2011.

\bibitem{Rajput2011}
Gagan Rajput and KC~Sharma.
\newblock Colossal enhancement in thermoelectric efficiency of weakly coupled
  double quantum dot system.
\newblock {\em Journal of Applied Physics}, 110(11):113723, 2011.

\bibitem{Costi2010}
TA~Costi and V~Zlati{\'c}.
\newblock Thermoelectric transport through strongly correlated quantum dots.
\newblock {\em Physical Review B}, 81(23):235127, 2010.

\bibitem{Tsaousidou2010}
M~Tsaousidou and GP~Triberis.
\newblock Thermoelectric properties of a weakly coupled quantum dot: Enhanced
  thermoelectric efficiency.
\newblock {\em Journal of Physics: Condensed Matter}, 22(35):355304, 2010.

\bibitem{Tagani2012}
M~Bagheri Tagani and H~Rahimpour Soleimani.
\newblock Thermoelectric effects through weakly coupled double quantum dots.
\newblock {\em Physica B: Condensed Matter}, 407(4):765--769, 2012.

\bibitem{Wojcik2016}
Krzysztof~P W{\'o}jcik and Ireneusz Weymann.
\newblock Thermopower of strongly correlated t-shaped double quantum dots.
\newblock {\em Physical Review B}, 93(8):085428, 2016.

\bibitem{Donsa2014}
S~Donsa, S~Andergassen, and K~Held.
\newblock Double quantum dot as a minimal thermoelectric generator.
\newblock {\em Physical Review B}, 89(12):125103, 2014.

\bibitem{Zhang2007}
Zhi-Yong Zhang.
\newblock Thermopower of double quantum dots: Fano effect and competition
  between kondo and antiferromagnetic correlations.
\newblock {\em Journal of Physics: Condensed Matter}, 19(8):086214, 2007.

\bibitem{Wierzbicki2011}
M~Wierzbicki and R~Swirkowicz.
\newblock {Influence of interference effects on thermoelectric properties of
  double quantum dots}.
\newblock {\em Phys. Rev. B}, 84(7):075410, 2011.

\bibitem{Chen2000}
Xiaoshuang Chen, H.~Buhmann, and L.~W. Molenkamp.
\newblock Thermopower of the molecular state in a double quantum dot.
\newblock {\em Phys. Rev. B}, 61:16801--16806, 2000.

\bibitem{Trocha2012}
Piotr Trocha and J{\'{o}}zef Barna{\'{s}}.
\newblock {Large enhancement of thermoelectric effects in a double quantum dot
  system due to interference and Coulomb correlation phenomena}.
\newblock {\em Phys. Rev. B}, 85(8):085408, 2012.

\bibitem{VanderWiel2002}
Wilfred~G Van~der Wiel, Silvano De~Franceschi, Jeroen~M Elzerman, Toshimasa
  Fujisawa, Seigo Tarucha, and Leo~P Kouwenhoven.
\newblock Electron transport through double quantum dots.
\newblock {\em Reviews of Modern Physics}, 75(1):1, 2002.

\bibitem{zhang2018}
Yanchao Zhang, Zhimin Yang, Xin Zhang, Bihong Lin, Guoxing Lin, and Jincan
  Chen.
\newblock Coulomb-coupled quantum-dot thermal transistors.
\newblock {\em EPL (Europhysics Letters)}, 122(1):17002, 2018.

\bibitem{Galpin2006}
Martin~R Galpin, David~E Logan, and Hulikal~Ramaiyengar Krishnamurthy.
\newblock Renormalization group study of capacitively coupled double quantum
  dots.
\newblock {\em Journal of Physics: Condensed Matter}, 18(29):6545, 2006.

\bibitem{Ferreira2011}
Irisnei~L Ferreira, PA~Orellana, GB~Martins, FM~Souza, and E~Vernek.
\newblock Capacitively coupled double quantum dot system in the kondo regime.
\newblock {\em Physical Review B}, 84(20):205320, 2011.

\bibitem{Purkayastha2015}
Archak Purkayastha, Abhishek Dhar, and Manas Kulkarni.
\newblock {Out of equilibrium open quantum systems: a comparison of approximate
  Quantum Master Equation approaches with exact results}.
\newblock {\em Phys. Rev. A - At. Mol. Opt. Phys.}, 93(6), 2015.

\bibitem{PhysRevB.82.235307}
S.~Koller, M.~Grifoni, M.~Leijnse, and M.~R. Wegewijs.
\newblock Density-operator approaches to transport through interacting quantum
  dots: Simplifications in fourth-order perturbation theory.
\newblock {\em Phys. Rev. B}, 82:235307, 2010.

\bibitem{MeirWingreenFor}
Yigal Meir and Ned~S. Wingreen.
\newblock Landauer formula for the current through an interacting electron
  region.
\newblock {\em Phys. Rev. Lett.}, 68:2512--2515, 1992.

\bibitem{Schaller2014}
Gernot Schaller.
\newblock {\em {Open Quantum Systems Far from Equilibrium}}, volume 881 of {\em
  Lecture Notes in Physics}.
\newblock Springer International Publishing, Cham, 2014.

\bibitem{breuer2002theory}
H.P. Breuer and F.~Petruccione.
\newblock {\em The Theory of Open Quantum Systems}.
\newblock Oxford University Press, 2002.

\bibitem{bruus2004many}
Henrik Bruus and Karsten Flensberg.
\newblock {\em Many-body quantum theory in condensed matter physics: an
  introduction}.
\newblock Oxford University Press, 2004.

\bibitem{Gurvitz1996}
S.~A. Gurvitz and Ya.~S. Prager.
\newblock {Microscopic derivation of rate equations for quantum transport}.
\newblock {\em Phys. Rev. B}, 53(23):15932--15943, 1996.

\bibitem{CartesanTim2008}
Carsten Timm.
\newblock Tunneling through molecules and quantum dots: Master-equation
  approaches.
\newblock {\em Phys. Rev. B}, 77(19):195416, 2008.

\bibitem{Yoni_NDC}
Bingqian Xu and Yonatan Dubi.
\newblock Negative differential conductance in molecular junctions: an overview
  of experiment and theory.
\newblock {\em Journal of Physics: Condensed Matter}, 27(26):263202, 2015.

\bibitem{PhysRevB.76.035432}
Bhaskaran Muralidharan and Supriyo Datta.
\newblock Generic model for current collapse in spin-blockaded transport.
\newblock {\em Phys. Rev. B}, 76:035432, 2007.

\bibitem{Sierra2014}
Miguel~A. Sierra and David S{\'{a}}nchez.
\newblock {Strongly nonlinear thermovoltage and heat dissipation in interacting
  quantum dots}.
\newblock {\em Phys. Rev. B}, 90(11):115313, 2014.

\bibitem{belzig2005}
W.~Belzig.
\newblock Full counting statistics of super-poissonian shot noise in multilevel
  quantum dots.
\newblock {\em Phys. Rev. B}, 71:161301, Apr 2005.

\bibitem{PhysRevB.78.161406}
Padraig Murphy, Subroto Mukerjee, and Joel Moore.
\newblock Optimal thermoelectric figure of merit of a molecular junction.
\newblock {\em Phys. Rev. B}, 78:161406, 2008.

\bibitem{PhysRevB.44.1646}
C.~W.~J. Beenakker.
\newblock Theory of coulomb-blockade oscillations in the conductance of a
  quantum dot.
\newblock {\em Phys. Rev. B}, 44:1646--1656, 1991.

\bibitem{eitan}
Giovanni Viola, Sourin Das, Eytan Grosfeld, and Ady Stern.
\newblock Thermoelectric probe for neutral edge modes in the fractional quantum
  hall regime.
\newblock {\em Phys. Rev. Lett.}, 109:146801, 2012.

\end{thebibliography}

\end{document}